\pdfoutput=1
\documentclass[11pt,twoside,a4paper,cmspaper,final,collab]{cms-tdr}

\begin{document}\cmsNoteHeader{ecal-tb-paper}

\newlength\cmsFigWidth
\ifthenelse{\boolean{cms@external}}{\setlength\cmsFigWidth{0.49\textwidth}}{\setlength\cmsFigWidth{0.65\textwidth}}
\ifthenelse{\boolean{cms@external}}{\providecommand{\cmsLeft}{upper\xspace}}{\providecommand{\cmsLeft}{left\xspace}}
\ifthenelse{\boolean{cms@external}}{\providecommand{\cmsRight}{lower\xspace}}{\providecommand{\cmsRight}{right\xspace}}
\providecommand{\textcolor}[2]{#2}

\newcommand{\LOne}{L$1$\xspace}
\newcommand{\LSTwo}{LS-$2$\xspace}
\newcommand{\LSThree}{LS-$3$\xspace}
\newcommand{\LSFour}{LS-$4$\xspace}
\newcommand{\RunOne}{Run-$1$\xspace}
\newcommand{\RunTwo}{Run-$2$\xspace}
\newcommand{\RunThree}{Run-$3$\xspace}
\newcommand{\RunFour}{Run-$4$\xspace}
\newcommand{\PhaseOne}{Phase-$1$\xspace}
\newcommand{\PhaseTwo}{Phase-$2$\xspace}
\newcommand{\Hgg}{$\text{H}\to\gamma\gamma$\xspace}
\newcommand{\aeta}{\abs{\eta}}
\newcommand{\celsius}{\ensuremath{\,^\circ\text{C}}\xspace}
\newcommand{\ns}{\ensuremath{\,\text{ns}}\xspace}
\newcommand{\ps}{\ensuremath{\,\text{ps}}\xspace}

\title{Performance of the front-end electronics of the CMS electromagnetic calorimeter barrel for the High-Luminosity LHC}

\author*{The CMS electromagnetic calorimeter group}

\date{\today}

\abstract{
The performance of the CMS electromagnetic calorimeter upgraded readout electronics, developed for the High-Luminosity phase of the LHC, is discussed. Data collected in two beam test campaigns conducted in 2018 and 2021 at the H4 and H2 beam lines of the CERN SPS are analyzed. Time and energy resolutions are measured on a $5\times 5$ matrix of lead tungstate crystals equipped with prototypes of the new front end readout electronics, using electron and pion beams of energies spanning from 25 to 250\GeV. In both campaigns the constant term of the energy resolution is measured to be better than 0.6\% and the time resolution for electrons with energies above 50\GeV is measured to be better than 30\ps, fulfilling the design requirements.
}


\author[1,2]{The CMS electromagnetic calorimeter group\note{Corresponding authors: Chiara Basile chiara.basile@cern.ch, Federico Ferri federico.ferri@cern.ch.}\note{Complete author list at the end of the document.}}

\hypersetup{%
pdfauthor={The CMS electromagnetic calorimeter group},%
pdftitle={Performance of the front-end electronics of the CMS Electromagnetic Calorimeter barrel for the High-Luminosity LHC},%
pdfsubject={ECAL},%
pdfkeywords={CMS ECAL, electronics, calorimetry, HL-LHC}}

\maketitle

\section{Introduction}

The Compact Muon Solenoid (CMS) apparatus~\cite{CMS:Detector-2008,CMS:PRF-21-001} at the CERN Large Hadron Collider (LHC) is a multipurpose, nearly hermetic detector, designed to trigger on~\cite{CMS:TRG-17-001,CMS:TRG-12-001,CMS:TRG-19-001} and identify electrons, muons, photons, and (charged and neutral) hadrons~\cite{CMS:EGM-17-001,CMS:MUO-16-001,CMS:TRK-11-001}. Its central feature is a superconducting solenoid of 6\unit{m} internal diameter and 12.5\unit{m} length, providing a magnetic field of 3.8\unit{T}. Within the solenoid volume are a silicon pixel and strip tracker, a lead tungstate crystal electromagnetic calorimeter (ECAL), and a brass and scintillator hadron calorimeter (HCAL), each composed of a barrel and two endcap sections. Forward calorimeters extend the pseudorapidity coverage provided by the barrel and endcap detectors. Muons are reconstructed using gas-ionization detectors embedded in the steel flux-return yoke outside the solenoid. More detailed descriptions of the CMS detector, together with a definition of the coordinate system used and the relevant kinematic variables, can be found in Refs.~\cite{CMS:Detector-2008,CMS:PRF-21-001}.

\begin{figure}
        \centering
        \includegraphics[width=.75\textwidth]{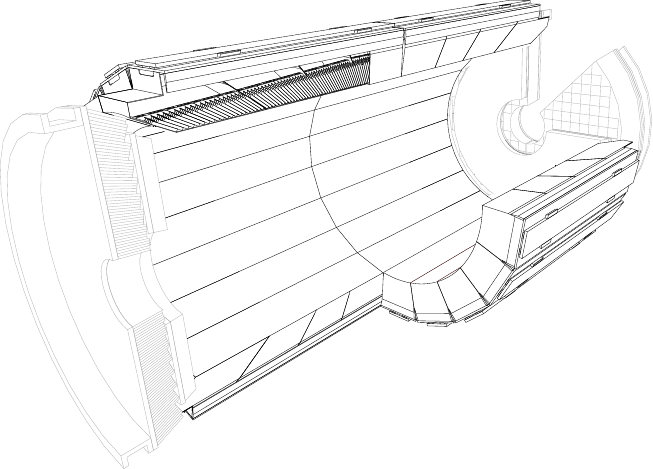} \caption{A three dimensional view of the ECAL. In the barrel section, the cutaway views illustrate the disposition of modules within a supermodule and of crystals within a module. The endcaps are shown in grey as they will be replaced by a new subsystem as part of the upgrade for the high-luminosity LHC.}
        \label{fig:ecal}
\end{figure}

The CMS ECAL is a homogeneous and hermetic electromagnetic calorimeter made of lead tungstate (PbWO$_4$) scintillating crystals. The high density ($8.28\unit{g/cm}^{3}$), short radiation length (8.9\mm), and small Moli\`ere radius (22\mm) of the PbWO$_4$ enabled the construction of a compact calorimeter with fine lateral granularity. The ECAL comprises a central barrel and two endcaps. The barrel, about 6\unit{m} long and with an inner radius of about 1.3\unit{m}, covers the pseudorapidity region $\aeta < 1.48$. It consists of 36 elements, known as supermodules (SM), arranged in two rings of 18, each covering half the length of the barrel. Each SM covers a range of $20\de$ in azimuthal angle ($\varphi$), as shown in the schematic drawing of figure~\ref{fig:ecal}.
The $61\,200$ crystals in the barrel are arranged in a quasi-projective geometry, oriented toward the nominal interaction point, with their axis tilted by approximately $3\de$ in both $\eta$ and $\varphi$. The crystals come in 17 different shapes, depending on their position along $\eta$. In the barrel, the emitted light is detected with avalanche photodiodes (APDs) operating at a gain of 50, corresponding to a bias voltage between 350 and 420\unit{V}. A more detailed description of the ECAL, including the endcaps, can be found in Ref.~\cite{CMS:1997ema}.

The ECAL was designed to achieve an excellent energy resolution up to an integrated luminosity of 500\fbinv over ten years of data taking at a peak luminosity of $1\times 10^{34}\percms$. The performance of the LHC during \RunTwo and \RunThree has exceeded the peak luminosity by more than a factor of two, with the ECAL continuing to deliver excellent results~\cite{CMS:PRF-21-001,CMS:EGM-18-002}. The High-Luminosity (HL) phase of the LHC, also called \PhaseTwo of the LHC, is expected to deliver an integrated luminosity of 3000\fbinv with radiation levels in the detector six times higher than the nominal LHC design~\cite{Aberle:2749422}. To maintain the current performance in such a difficult environment the ECAL barrel must be upgraded~\cite{CMS:TDR-015}. The ECAL endcap, along with the entire CMS forward calorimetry, will be completely replaced by a silicon-based high-granularity calorimeter~\cite{CERN-LHCC-2017-023}.

This paper presents the results of studies of the performance of the upgraded electronics for the ECAL, made with a $5\times 5$ matrix of $\text{PbWO}_4$ crystals and high energy electron or pion beams. The data were collected in separate campaigns, in 2018 and 2021, in the H4 and H2 test beams at the CERN SPS. In the following, the current electronics is referred to as the ``legacy electronics'' and the upgraded version as the ``\PhaseTwo electronics''.
Section~\ref{sec:hl:upgrade} generally describes the ECAL barrel upgrade for HL-LHC, and details the components of the \PhaseTwo electronics. Section~\ref{sec:tb:campaigns} describes the experimental setup of the beam test campaigns. The strategy and the results for the signal reconstruction are detailed in section~\ref{sec:tb:timereference}, \ref{sec:tb:reco}, and \ref{sec:xtal_intercalib}.
Finally, the results concerning time and energy resolution measurements are presented in section~\ref{sec:tb:results}.

\section{The upgrade of the CMS electromagnetic calorimeter}
\label{sec:hl:upgrade}

The primary predicted effects of operating at HL-LHC luminosities include a decrease in the light transmission of the PbWO$_4$ crystals, caused by irreversible radiation damage of the crystal lattice, and an increase of the APD leakage current, due to bulk damage in the silicon. These effects contribute to a reduction in the signal to noise ratio.

Studies have been conducted to assess the impact of the radiation on crystals and photodetectors, demonstrating that both components can be successfully operated up to the end of HL-LHC~\cite{CERN-LHCC-2017-011}.
To mitigate the increase in the bulk current of the APDs, the operating temperature will be lowered from the current 18$\celsius$ to 9$\celsius$. This will not require modification to the existing on-detector cooling services.

To cope with higher trigger rates, improve the rejection of signals from direct ionization of the APDs (spikes), and mitigate the effect of additional interactions within the same or neighbouring bunch crossings (in-time and out-of-time pileup, respectively), the front-end and back-end ECAL barrel readout electronics will be replaced during the Long Shutdown 3 of the LHC, which is scheduled from 2026 to 2029.

\subsection{Readout electronics}
\label{sec:hl:electronics}

\begin{figure}
        \centering
        \includegraphics[width=.5\textwidth]{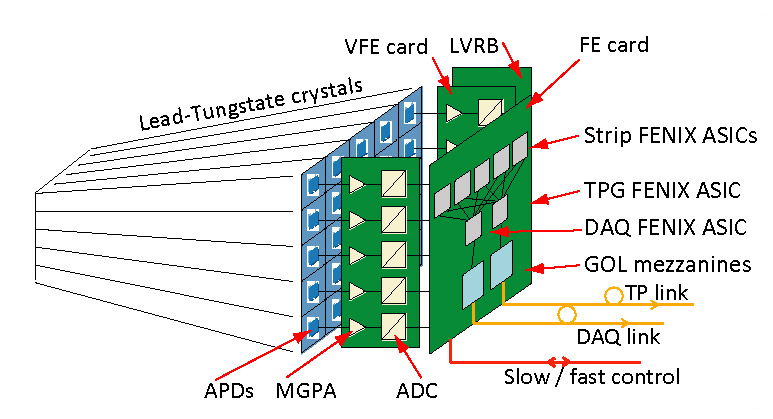}%
        \includegraphics[width=.5\textwidth]{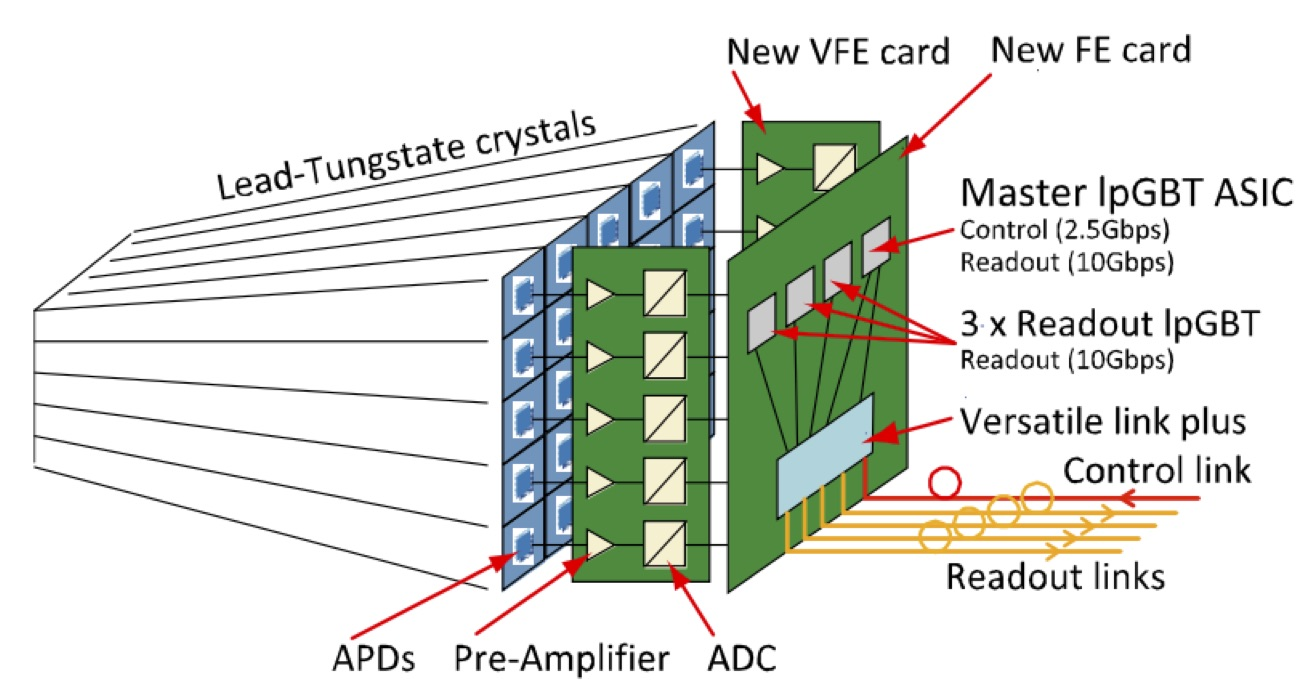}%
        \caption{Layout of the main components of a readout unit of the legacy electronics (left) and of the \PhaseTwo electronics  (right). The motherboards--not shown in here--are located parallel to the FE boards, between the crystals and the 5 VFEs. A low-voltage board is present in both systems, although shown only for the legacy one.}
        \label{fig:hl:ru}
\end{figure}

\begin{figure}
        \centering
        \includegraphics[width=\textwidth]{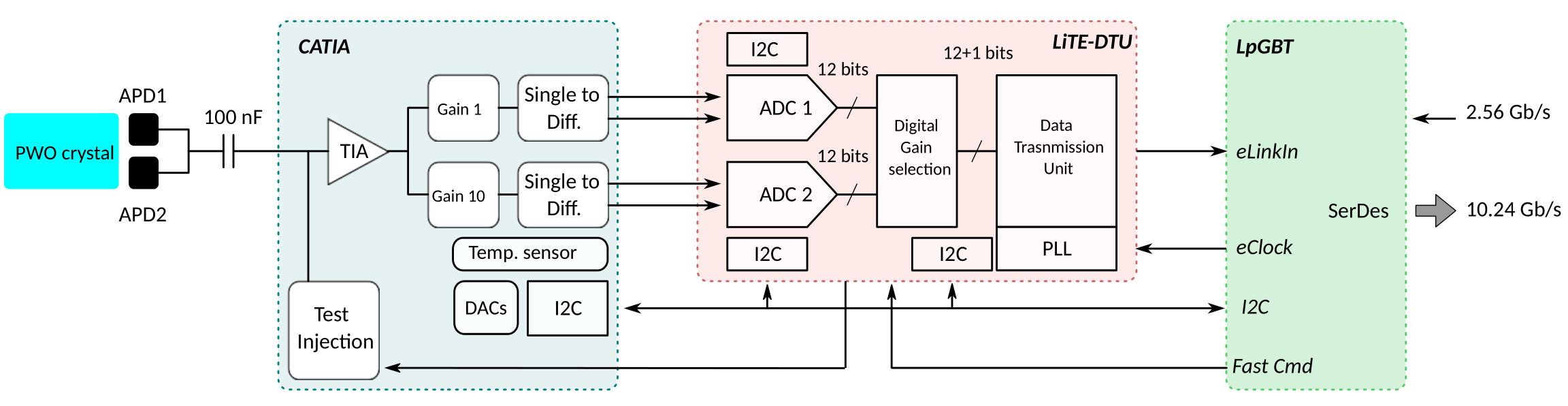}%
        \caption{Layout of the main components of the \PhaseTwo electronics along the signal path from the photodetectors to the back-end board. The CATIA and LiTE-DTU ASICs are located on the VFE board, while the green box represents the FE board.}
        \label{fig:hl:electronics}
\end{figure}

A schematic representation of both the legacy and \PhaseTwo electronics is shown in figure~\ref{fig:hl:ru}. The arrangement of an array of $5\times 5$ crystals forming a single readout unit is matched to the motherboards that distribute bias voltage to the APDs and low voltage to the Very Front End (VFE) and Front End (FE) cards. The motherboards contain only passive components and have been tested, demonstrating that they can operate throughout the entire HL-LHC, thus requiring no refurbishment. Consequently, the geometrical layout of the legacy electronics has been retained in the upgraded design.

\subsubsection{Front-end electronics}

One VFE card reads out five channels; for each channel, the \PhaseTwo VFE card will include an amplifier, a digitizer, and a data-transmission unit, as illustrated in figure~\ref{fig:hl:electronics}.
The amplifier ASIC, CATIA~\cite{Gevin:2020a6} (CAlorimeter TIA), is based on a trans-impedance amplifier (TIA) that generates a voltage representation of the photocurrent produced by the APD. Its accuracy is primarily determined by the system bandwidth, limited by the kapton connection between the APDs and the motherboard, which is approximately 35\unit{MHz}. The bandwidth of the CATIA alone reaches 50\unit{MHz}.
The output electronics pulses from the amplifier are approximately 40\unit{ns} long, compared to about 350\unit{ns} in the legacy system. The shorter pulse significantly reduces the overlap of signals from out-of-time pileup. Two different gains ($\times 1$ and $\times 10$) provide a dynamic range covering signals up to 2\TeV, with the gain switch occurring at around 200\GeV, with the precise value depending on the crystal light-yield.

The preamplifier is followed by a data conversion and transmission ASIC, the LiTE-DTU~\cite{LiTE-DTU} (Lisboa-Torino ECAL data transmission unit). Each LiTE-DTU receives the analog signals from both gains of the CATIA, digitizes them at 160\unit{MS/s}, using 12-bits ADCs, and transmits the data to the FE board.
To optimize the output data bandwidth, only the signal from the highest unsaturated gain is transmitted, and a lossless data compression scheme is applied. Whereby 6\unit{bits} are transmitted for baseline samples, and 13\unit{bits} for samples that exceed the 6-bit representation.
The gain switch is handled by sending an extendable window of samples in gain 1, positioned around the first saturated sample, ensuring that a pulse that saturates in gain 10 is fully acquired in gain 1.
The sampling frequency, 4 times higher than in the legacy system, has been chosen to achieve a time resolution of approximately 30\unit{ps} at energies higher than 50\GeV, which are in the relevant range for photons from Higgs boson to diphoton decays.

This sampling frequency provides performance that is independent of the sampling phase, and, along with the large bandwidth of the system, it also gives better discrimination between the signals from electromagnetic showers and the spikes. The latter can be suppressed almost completely by the \LOne trigger.
Additionally, thanks to the improved time resolution, the precise knowledge of the electromagnetic particle arrival time will allow for a better mitigation of the in-time pileup~\cite{CERN-LHCC-2017-011}.
The FE card utilizes recent developments in radiation-tolerant optical links, such as the Versatile Link plus~\cite{TROSKA2023168208} and the lpGBT~\cite{lpGBT_2024}, to serialize and transmit the digital data stream from each individual LiTE-DTU to the back-end electronic system for processing.

\subsubsection{Back-end electronics}

The back-end electronics transmits the data to the central data acquisition (DAQ) system of CMS and provides single-crystal trigger primitives to the \LOne trigger system. Hence, the ECAL information available at \LOne will be $25$ times greater than that provided by the legacy system, which is based on a $5\times 5$ channel readout unit.
The Barrel Calorimeter Processing boards~\cite{Loukas:2644903} are designed with commercially available FPGAs and high speed optical links. The FPGAs are sufficiently powerful to provide a rejection of the spikes based both on the pulse shape and on topological variables.

\section{Beam test campaigns}
\label{sec:tb:campaigns}

The first prototypes of the ECAL \PhaseTwo electronics have been tested at the CERN SPS North Area, on the H$4$ and H$2$ beam lines~\cite{H4H2_sps}. These tests concentrated primarily on the CATIA, LiTE-DTU, and low-voltage regulator boards. They were conducted in stages, starting with initial discrete-component prototypes and progressing to nearly final versions of the ASICs. In the absence of a complete set of the prototype components of the CMS DAQ chain, additional \emph{ad-hoc} boards with similar functionalities have been produced to read out the data and to provide precise clock and trigger signals.

This paper focuses on the results obtained from 2018 and 2021 beam test campaigns, featuring the latest available prototypes at the time of the tests.

\subsection{Beamline setup}

The H4 beam is supplied with particles produced by 400 GeV protons extracted from the CERN SPS striking a 30\cm long beryllium target, and can be tuned to provide pions, electrons or photons.
Secondary beams of charged pions are extracted directly after the primary target, achieving a purity greater than $90\%$, with the main contamination arising from the presence of electrons and muons.
Tertiary beams of electrons or positrons are obtained by the conversion of photons from the decays of secondary neutral pions. The beam purity depends on the energy and is better than $98\%$. A specific set of beam optics provides a relative momentum spread $\Delta p/p$ for electrons and positrons as low as $0.5\%$~\cite{H4H2_sps}.

\begin{figure}[t]
        \centering
        \includegraphics[width=\textwidth]{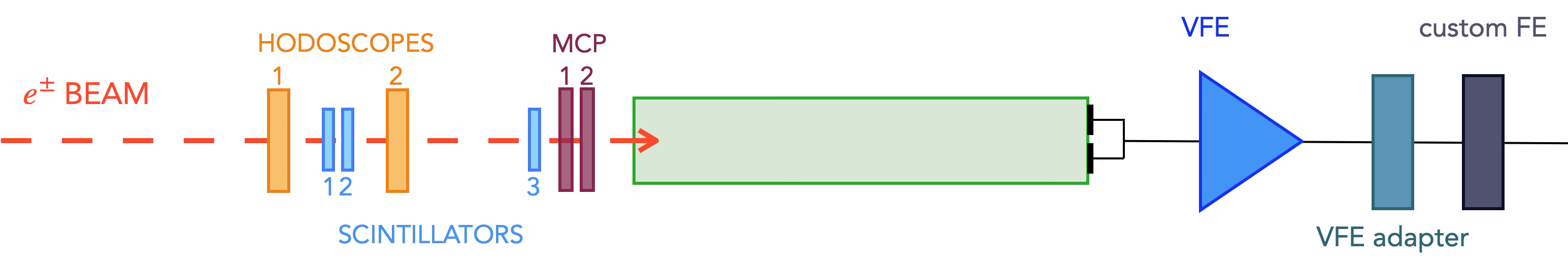}%
        \caption{Schematic representation of the experimental setup along the H4 beamline.}
        \label{fig:tb:setup}
\end{figure}

Two sets of hodoscopes are located in the beamline upstream of the ECAL crystals, each consisting of two planes of orthogonal scintillating fibers. The hodoscopes provide a reference position in the transverse plane with a spatial resolution of approximately 150\mum. A reference for the arrival time of the beam particles is provided by two microchannel plate detectors (MCPs)~\cite{Barnyakov_2018}, with a resolution of approximately 15\unit{ps}. A schematic representation of the experimental setup in 2018 and 2021 is shown in figure~\ref{fig:tb:setup}.

The crystals are arranged in a $5\times 5$ matrix, emulating an ECAL readout unit in the second module ($\aeta \approx 1.09$), including the cooling system of the readout electronics. At variance with the arrangement in the ECAL, where all crystals at a given $\eta$ are of the same type, the matrix hosts crystals of type 5, 6, and 7. The matrix is installed inside a copper case, water-cooled with a temperature stability of 0.1\celsius, in line with the nominal detector operations of CMS. The APDs are connected to the input of the VFE boards using Kapton-insulated cables of type and length identical to those used in CMS.

\begin{figure}
        \centering
        \begin{minipage}{0.40\textwidth}
                \subfigure[]{\includegraphics[width=\textwidth]{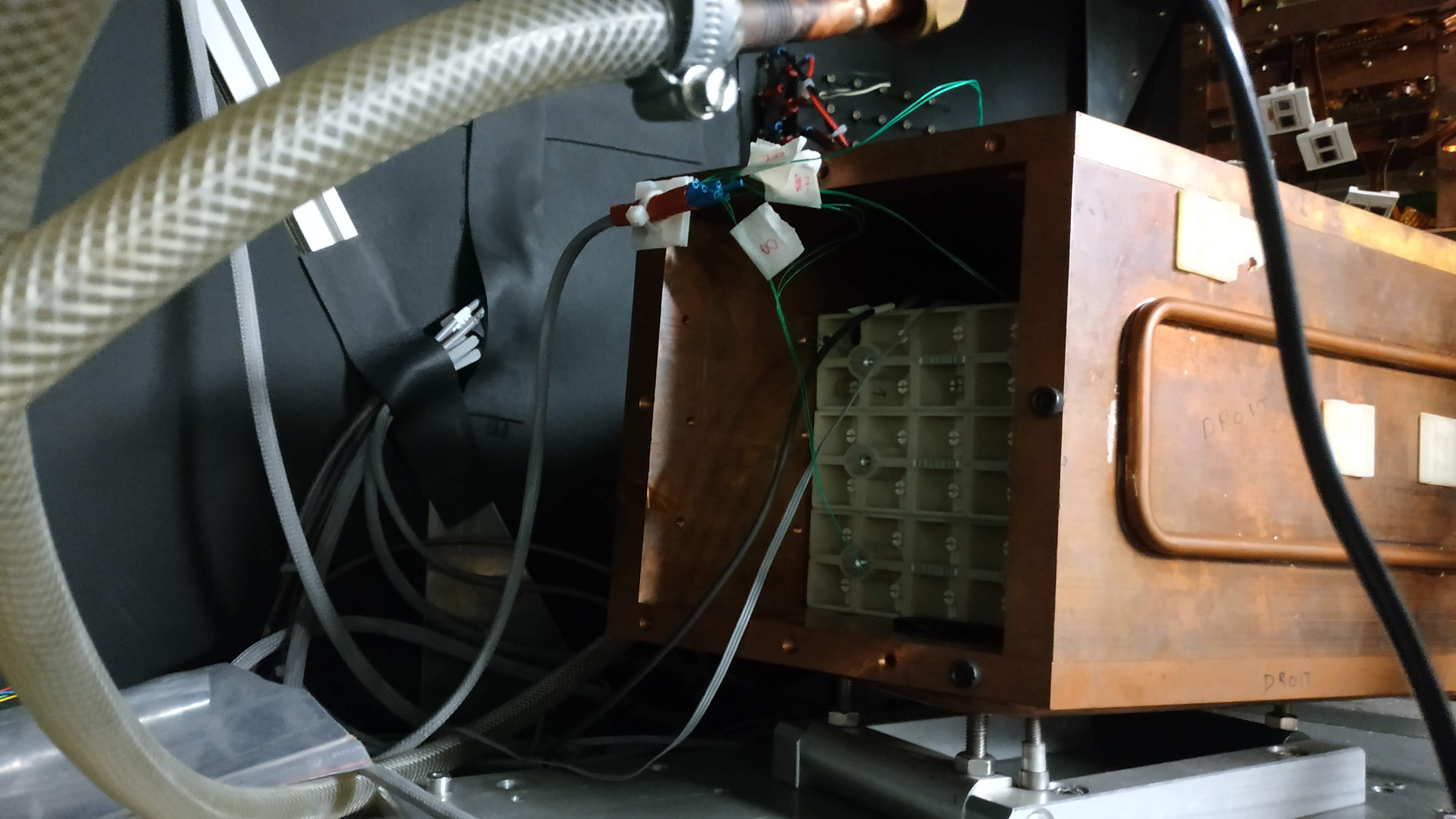}}\\
                \subfigure[]{\includegraphics[width=\textwidth]{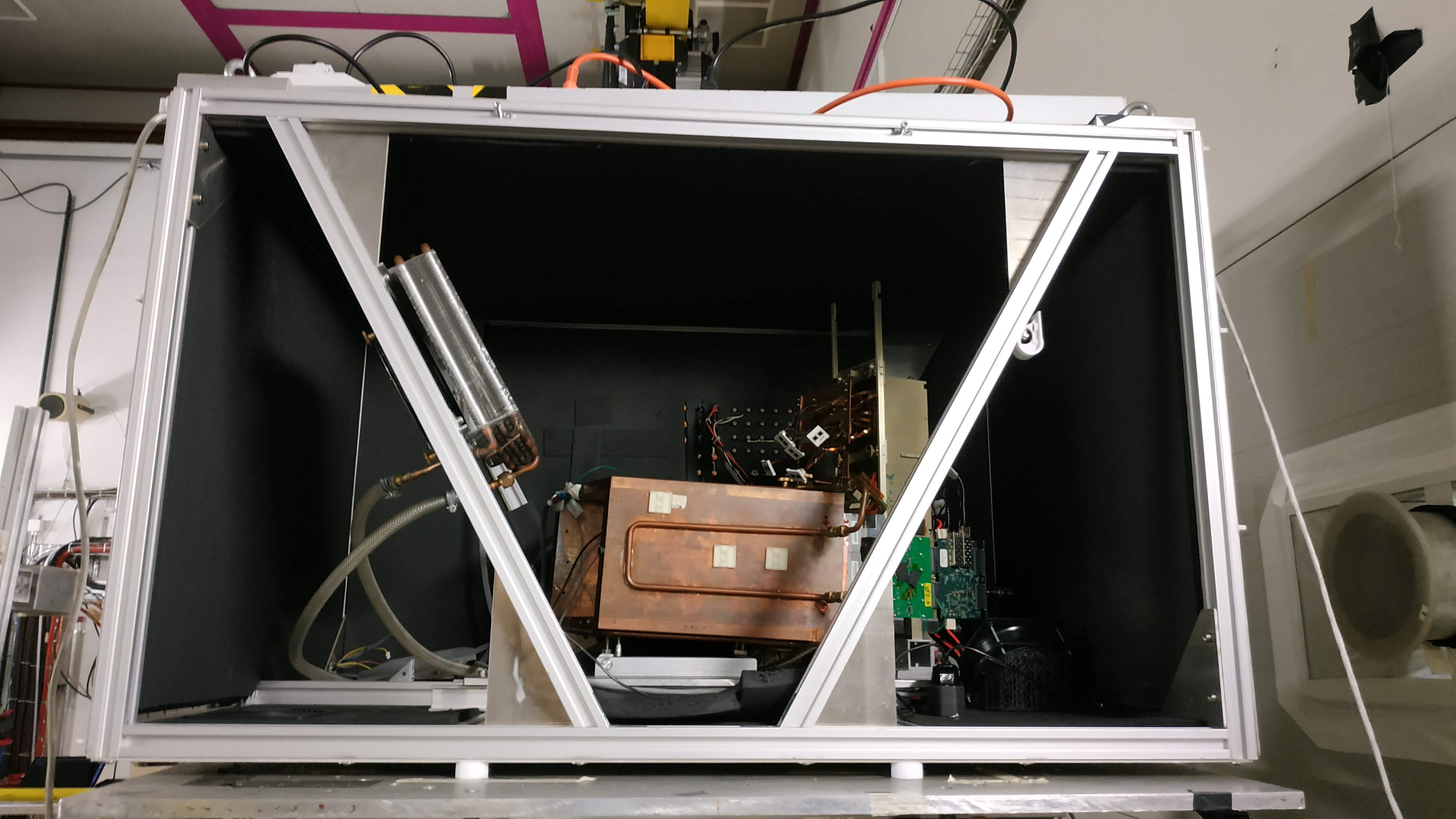}}
        \end{minipage}
        \begin{minipage}{0.3675\textwidth}
                \subfigure[]{\includegraphics[width=\textwidth]{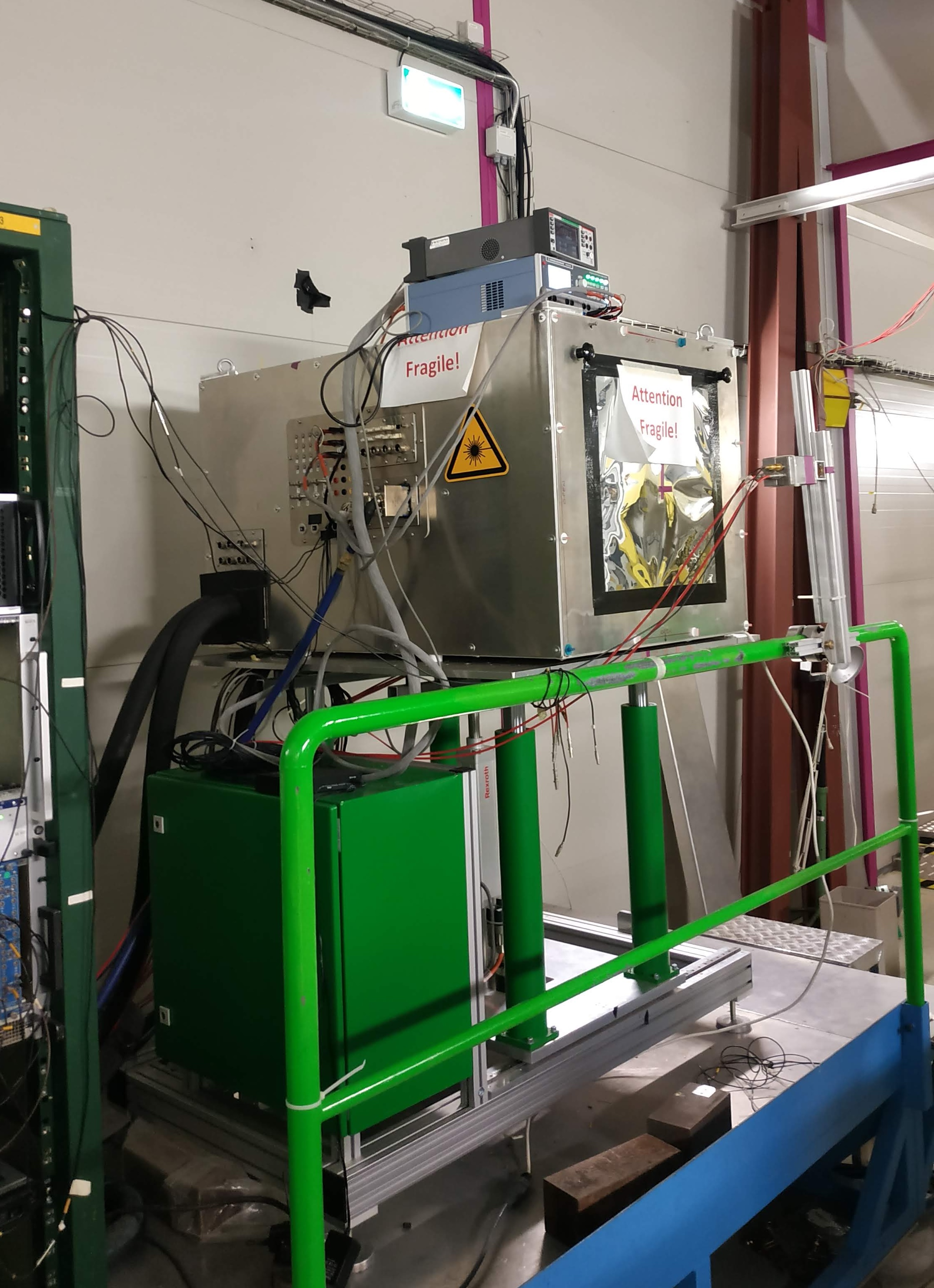}}
        \end{minipage} 
        \caption{The housing for the crystal array, visible within the copper case (a). The aluminum box containing the copper case (b). Also visible, above the case towards the rear, are unmounted APDs, used to measure with pions the signal produced by direct ionization. On the right side of the box, at the back, the patch panel for the VFE boards can be seen. The aluminum box is mounted on a remotely controlled table that allows the array to be moved in the transverse plane to center any selected crystal in the beam (c).}
        \label{fig:tb:coldbox}
\end{figure}

The copper case and the readout electronics are housed within a larger light-tight, thermally insulated aluminium box, which also contains an air convection system along with temperature and humidity sensors. A $20\times 20\cm^2$ carbon fiber window in the front panel of the box reduces the amount of material encountered by the beam.
The temperature of the entire area around the box is kept constant at 18$\celsius$. The temperature of the copper case can be adjusted to the operating conditions of the legacy system (18$\celsius$) and of the \PhaseTwo system (9$\celsius$). The dew point inside the box, regulated using a dry air flow, has been continuously monitored. The box is mounted on a remotely controllable table that can be moved in the transverse plane allowing the beam to impinge on each crystal of the matrix, in a way that replicates the geometry of the ECAL in CMS with respect to particles originating from the nominal interaction point. Pictures of the matrix housing are shown in figure~\ref{fig:tb:coldbox}.

\subsection{Front-end electronics prototypes}

The 2018 campaign was conducted with VFEs incorporating a first prototype version (v0) of the CATIA amplifier. The objective of implementing this purely analogue device was to validate the TIA design, and it was paired to a 14-bit commercial ADC Analog-Device AD9642 digitizing signals at 160\unit{MS/s}. The VFEs were read out by an \emph{ad-hoc} board designed to support a data stream from a single CATIA gain.
The first complete prototype of a VFE, including the ASICs CATIA v1.2 and LiTE-DTUs v1.2, was tested for the first time during the 2021 campaign. The CATIA v1.2 prototype retains a TIA block nearly identical to that of the initial CATIA v0 prototype, with the addition of digital control, a differential output stage, and a pulse injection system. The LiTE-DTU v1.2 prototype includes all the final features but an improved channel-to-channel synchronization and bit alignment.

\subsection{Trigger and data acquisition}

The trigger to the DAQ readout is given by the coincidence of signals from at least two of three plastic scintillators measuring $3\times 3\cm^2$, $1\times 1\cm^2$ and $2\times 2\cm^2$, as in figure~\ref{fig:tb:setup}, and aligned along the beamline to select a portion of the beam that covers the desired surface.

The DAQ software is based on a modular architecture that provides for the integration of several detector components and is capable of providing a run controller, an event builder, a fast data quality monitoring, and a system for data storage and bookkeeping. A more complete description of the DAQ can be found in~\cite{Marini:2018twl}.

\section{Time reference}
\label{sec:tb:timereference}

The MCPs along the beamline provide a precise time reference for the arrival time of the particles. The signal from the MCPs is digitized by means of a \textsc{caen v1742} VME board, operating at a sampling frequency of 5\unit{GS/s}.
The \textsc{caen v1742} board is based on the \textsc{DRS4} ASIC, which processes the input signal through a switched-capacitor array of $1024$ cells~\cite{RITT2010486}. Each cell samples the signal with a delay of approximately 200\unit{ps} with respect to the previous one. This delay is fixed during the manufacture of the ASIC and has been calibrated using a precise sine wave generator.
Approximately $10\,000$ input sinusoidal signals of the same frequency were digitized, and a fit to these data was performed with the overall frequency and the delay of each cell treated as free parameters. This procedure provided an optimal synchronization.
The time jitter between input channels within each of the two blocks of eight inputs of the \textsc{caen v1742} board is negligible compared to the MCPs resolution.

The ECAL readout electronics and the \textsc{v1742} board utilize two different clocks to digitise respectively the APD (160\unit{MS/s}) and the MCP signals (5\unit{GS/s}). In order to measure the phase between the two clocks, the one used by the first is digitized with the \textsc{v1742} board, using a second channel. For each event, the relative phase $\bar t_{\textsc{ecal}} - \bar t_{\textsc{mcp}}$ between the particle arrival time, measured in the ECAL and in the MCP, within one ADC clock period ($T_\text{clk}$), can be defined as:
\begin{align}
        \Delta t &\equiv \bar t_{\textsc{mcp}} - t_{clk}\\
        \bar t_{\textsc{ecal}} - \bar t_{\textsc{mcp}} &\equiv  (\bar t_{\text{\textsc{ecal}}} - \Delta t) \ \bmod T_\text{clk}
        \label{form:tb:tphase}
\end{align}
where the times $\bar t_{\text{X}}$ and $t_{\text{clk}}$ are all referred to the start of the acquisition windows in the corresponding boards. The time $t_{\text{clk}}$ is referred to the rising edge of the clock signal digitized by the \textsc{v1742}, and $T_\text{clk} = 6.238\ns$.

The quantities $\bar t_{\textsc{ecal}}$ and  $\bar t_{\textsc{mcp}}$ are the reconstructed time of the signal maximum, measured separately in the ECAL and the MCPs.

\section{Signal amplitude and time reconstruction}
\label{sec:tb:reco}

The amplitude $\mathcal{A}$ and the time $\bar t$ of the signals from the ECAL and the MCPs are extracted by fitting the measured waveforms with two template shapes. In the fit, a $\chi^2$ is minimized, defined as
\begin{equation}
        \chi^2 = \sum_i \frac{(A(t_i) - \mathcal{A}\cdot\mathcal{S}(t_i - \bar t))^2}{\sigma_{\text{N}}^2},
\end{equation}
where $A(t_i)$ is the sample amplitude at the i-th sample time $t_i$, $\mathcal{S}$ is the normalized template shape evaluated at $t_i$, and $\mathcal{A}$ and $\bar t$ are the free parameters of the fit, representing the signal amplitude and arrival time of the pulse, respectively. The ECAL and MCPs signals are reconstructed fitting 9 and 40 points, respectively, around the maximum amplitude of the pulse.

By exploiting the time distribution of the particles inside a de-bunched extracted beam, which is random with respect to the DAQ clock phase, template shapes $\mathcal{S}$ for ECAL and MCPs pulses can be obtained by averaging normalized, phase-aligned signals.
\begin{figure}
        \centering
        \includegraphics[width=.55\textwidth]{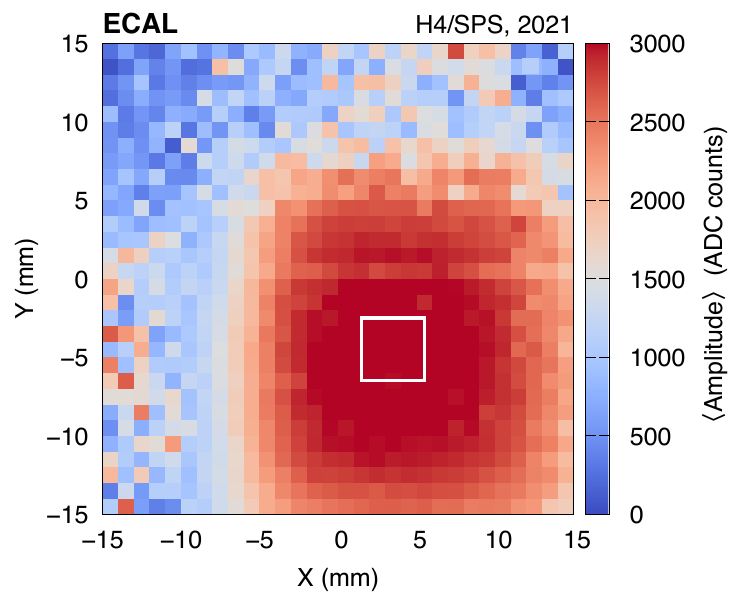}%
        \caption{Hodoscope heat map, where the $x$ and $y$ axis correspond to the coordinates on the hodoscope plates, and the $z$-axis represents the average signal amplitude measured in the crystal impacted by the electron beam. The white square contains the maximum shower energy at the position shown and indicates the region selected for the analysis.}
        \label{fig:hodo}
\end{figure}
For the ECAL, only signals from crystals directly impacted by the beam are used. Events are selected within a $4\times 4\mm^2$ window around the point of maximum shower containment, using the information from the hodoscopes, as shown in figure~\ref{fig:hodo}. Within this window, the variation of the average signal amplitude along $x$ and $y$ is below 1\%.
Additionally, to minimize the contribution from the pion contamination in the electron/positron beam, events are required to be associated with a signal of amplitude larger than 200\unit{ADC} counts in the MCP located furthest upstream.

In each event the phase and amplitude of the signal are initially reconstructed and then normalized and aligned. The first step is to increase the number of samples by a large amount, specifically by a factor of 50. This is performed in the frequency domain, using a Discrete Fourier Transformation (DFT) of the sampled signals, and by adding extra samples with values set at 0 in the frequency domain. An inverse DFT is then performed to obtain the oversampled signal. The oversampling provides a better resolution in the time domain, and gives a more precise alignment of the signals.
The phase, $\bar{t}_\text{o.p.}$, and the amplitude of the oversampled pulses are obtained by fitting five samples centred around the maximum one with a second order polynomial.
\begin{figure}
        \centering
        \includegraphics[width=1.\textwidth]{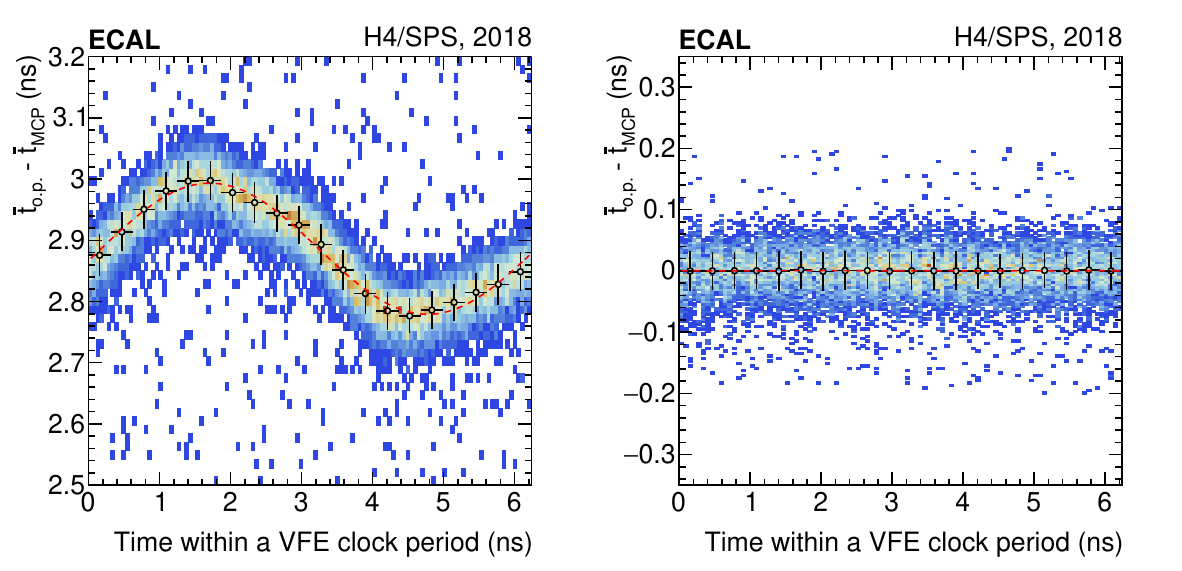}%
        \caption{Bias in alignment of ECAL pulses for the derivation of the template shape, $\bar{t}_\text{o.p.} - \bar{t}_{\textsc{mcp}}$, as a function of the ECAL sampling phase. Each point represents the mean of a Gaussian fit to the corresponding slice in $x$, while the error bars indicate the standard deviation of the Gaussian fit. A sinusoidal curve, shown in red, is fitted to the data and used to correct the bias. The right plot shows the residual bias after this correction has been applied.}
        \label{fig:tb:phasebias}
\end{figure}
This procedure leaves a bias in the reconstruction of the phase as a function of the sampling phase, as can be seen in figure~\ref{fig:tb:phasebias} left. This bias is corrected by applying a sinusoidal curve fitted to the data, and the result is shown in figure~\ref{fig:tb:phasebias} right.

\begin{figure}
        \centering
        \includegraphics[width=.5\textwidth]{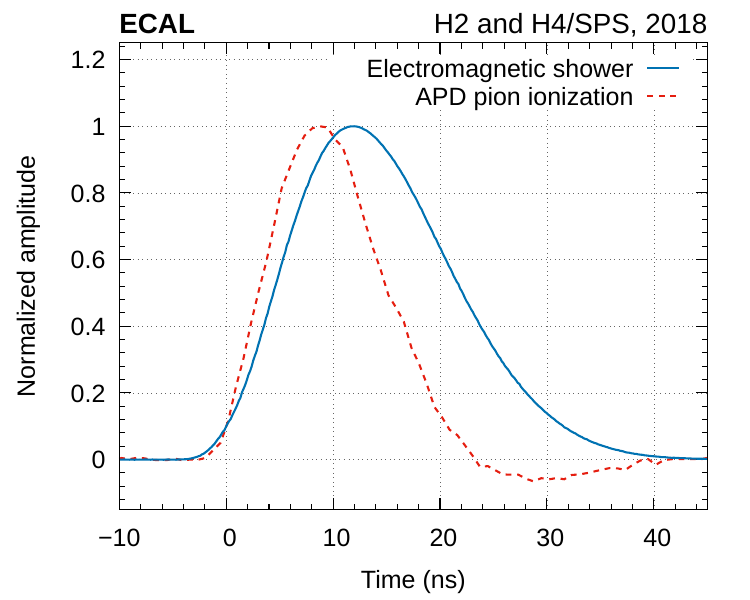}%
        \caption{Comparison of the signal shape for electromagnetic signal resulting from 150\GeV electrons (blue line) and for signals from direct ionization of the APDs measured with high energy pions (red line). The different smoothness of the curves is related to the different numbers of events available for deriving the shapes.}
\label{fig:tb:shapes}
\end{figure}
For the 2018 campaign, signal templates were derived for ECAL channels using electrons with an energy of 150\GeV. The choice of this energy is based on the larger available data sample, as all crystals have data at this energy, obtained when intercalibrating their response. Additionally, measurements made with a charged-pion beam in the H2 beamline allowed the derivation of template shapes for signals resulting from the direct ionization of the APD sensitive volume. The comparison, presented in figure~\ref{fig:tb:shapes}, reveals non-negligible differences in the pulse shapes, that can be exploited for their discrimination.

\begin{figure}[t]
        \begin{minipage}{0.5\textwidth}
                \centering
                \includegraphics[width=0.9\textwidth]{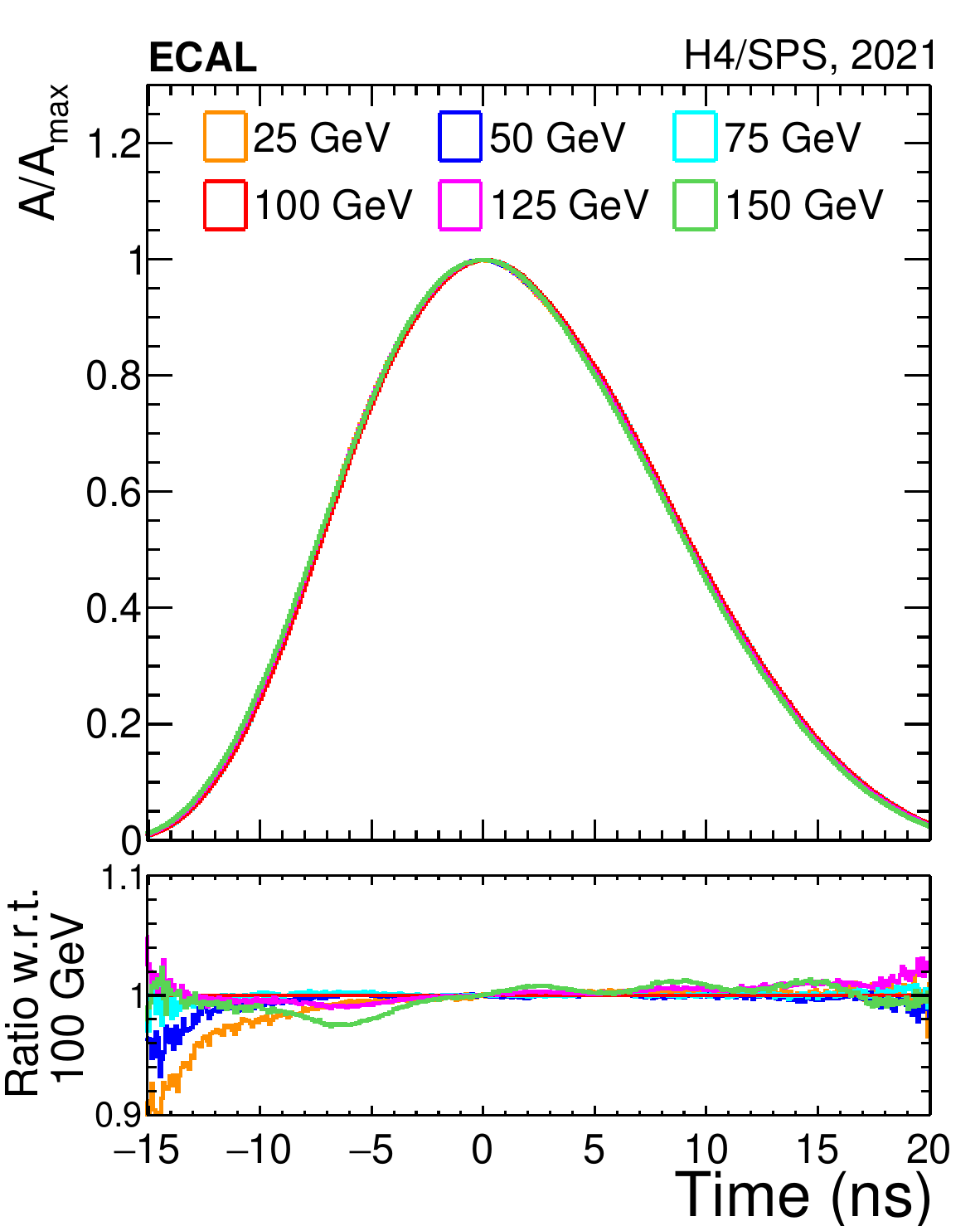}
        \label{fig:templatesVSenergy}
        \end{minipage}
        \begin{minipage}{0.5\textwidth}
                \centering
                \includegraphics[width=0.9\textwidth]{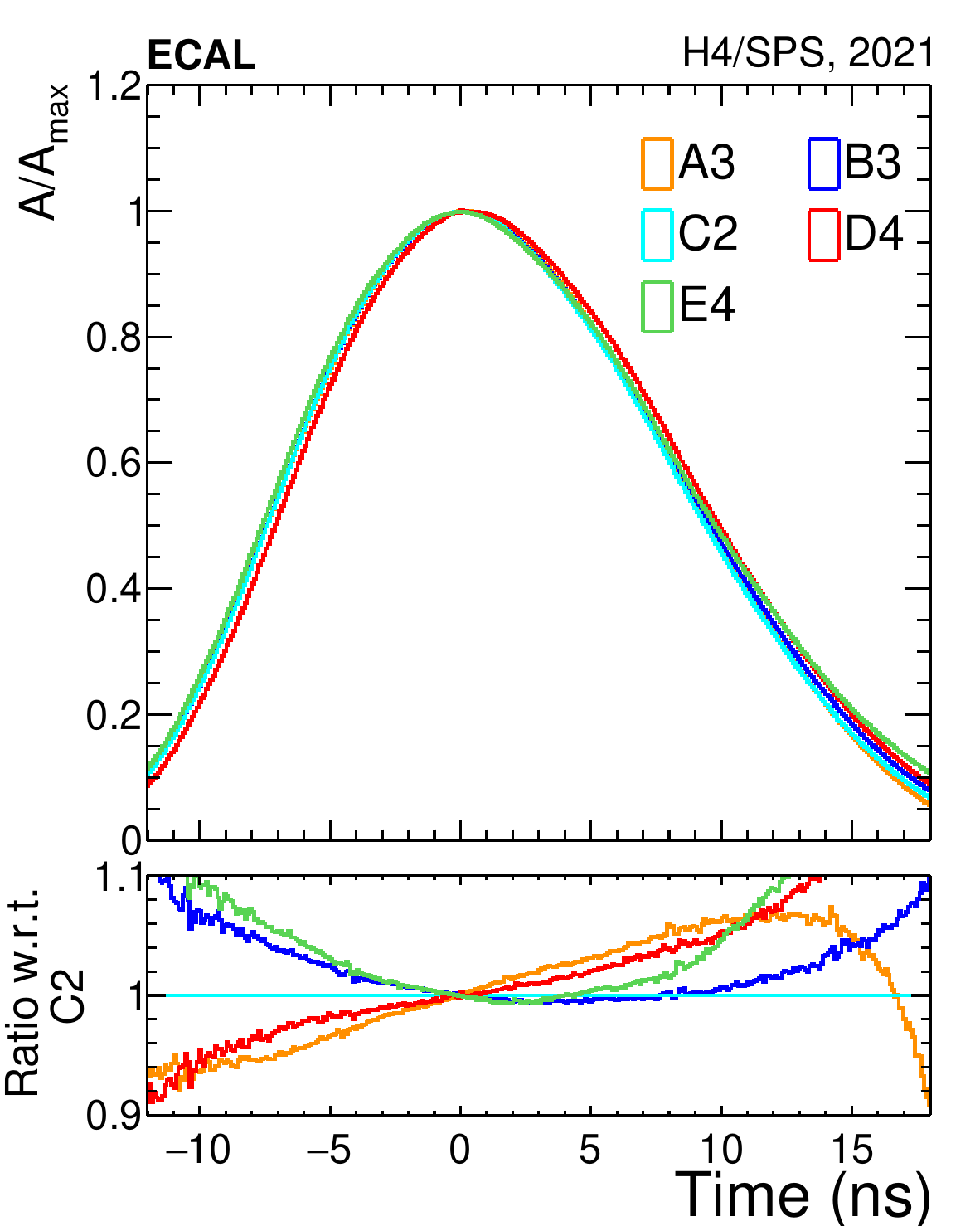}
        \label{fig:templatesVSvfes}
        \end{minipage} 
        \caption{Comparison of the signal shapes measured for a selected channel at different beam energies (left) and for different channels on different VFE boards (right). The signal shape for a given channel is independent of the electron energy, while different channels have different signal shapes.}
        \label{fig:templates}
\end{figure}

For the 2021 campaign, the dependency of the digitized waveforms on the particle energy and on the entire electronics readout chain, including the effects of light production and transmission in the crystals and the response of the APDs, is evaluated.
To investigate the dependence on the particle energy, templates have been derived for a single channel of the ECAL matrix adjacent to the central one and read out by the same VFE card. Electron beams with energies ranging from 25 to 150\GeV are used and the comparison, presented in the left plot of figure~\ref{fig:templates}, shows that the signal shapes are independent of the particle energy up to 125\GeV, i.e. within the energy range covered by gain 10. It was verified that, at a fixed beam energy, signal shapes for channels on the same VFE board are consistent.
To examine the dependence on the crystal and electronics readout chain, templates were derived for five crystals on different VFE boards, using electrons at a fixed energy of 100\GeV. The comparison, presented in figure~\ref{fig:templates} right, shows non-negligible differences in the pulse shapes, which need careful consideration when designing algorithms for amplitude and time reconstruction.

In light of these results, the signal shapes can be considered independent of particle energy up to 125\GeV, but they are specific to each individual channel. In the following analysis of 2021 data, amplitude and time are reconstructed using dedicated per-VFE templates derived at 100\GeV. For energies greater than 125\GeV, templates derived at 150\GeV are utilized.

\section{Channel intercalibration}
\label{sec:xtal_intercalib}

The energy intercalibration is obtained by comparing the response of individual channels to an electron beam directly impinging on each of them. Electrons of 150\GeV and of 100\GeV were used in the 2018 and 2021 beam tests, respectively. The intercalibration procedure is similar to that described in~\cite{Adzic:2008zza}.

\begin{figure}
        \centering
        \includegraphics[width=.49\textwidth]{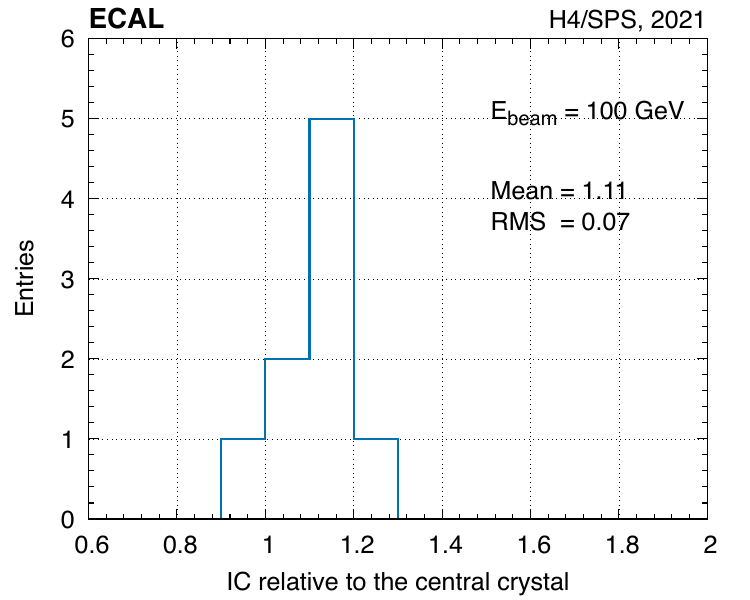}%
        \caption{Distribution of the values of the intercalibration coefficients relative to the central channel of the $3\times 3$ crystal matrix for the 2021 beam test. The spread of the coefficients is compatible with the expected spread of the crystal light yield.}
\label{fig:xtal_ic}
\end{figure}

Events for intercalibration are selected by requiring the signal amplitude in the MCP located upstream to be greater than 200 ADC, to veto events from pions. Additionally, by using the hodoscope information the impinging particle is required to be within a window of $4\times 4\mm^2$ centred on the position that maximizes the energy response of the crystal.
Figure \ref{fig:xtal_ic} shows the distribution of the intercalibration coefficients relative to the central crystal obtained in the 2021 beam test for the central $3\times 3$ matrix.

\section{Results}
\label{sec:tb:results}

During the 2018 campaign, measurements were made to compare the performances at operating temperatures of 18$\celsius$ and 9$\celsius$. The pulse shapes are found to be identical, and the light yield increases by about 18\% at the lower temperature, as expected from the known temperature dependence of $-2\%/\celsius$~\cite{CMS:TDR-015}. All subsequent measurements have therefore been made at a single temperature, chosen to be $18\celsius$ for ease of operation.

Although the energy resolution of the ECAL is driven by the physical design of the detector, which is unchanged, we have performed a cross-check measurement of the energy resolution using the energy deposited in a $3\times 3$ matrix centred around the crystal hit by the beam.

Events are selected using the same criteria applied for the intercalibration described in Section~\ref{sec:xtal_intercalib}. Results from the 2021 campaign are derived only for the high gain mode. The limited amount of data available with low gain precludes a reliable measurement of the energy resolution in this mode.

\begin{figure}[t]
        \centering
        \includegraphics[width=.5\textwidth]{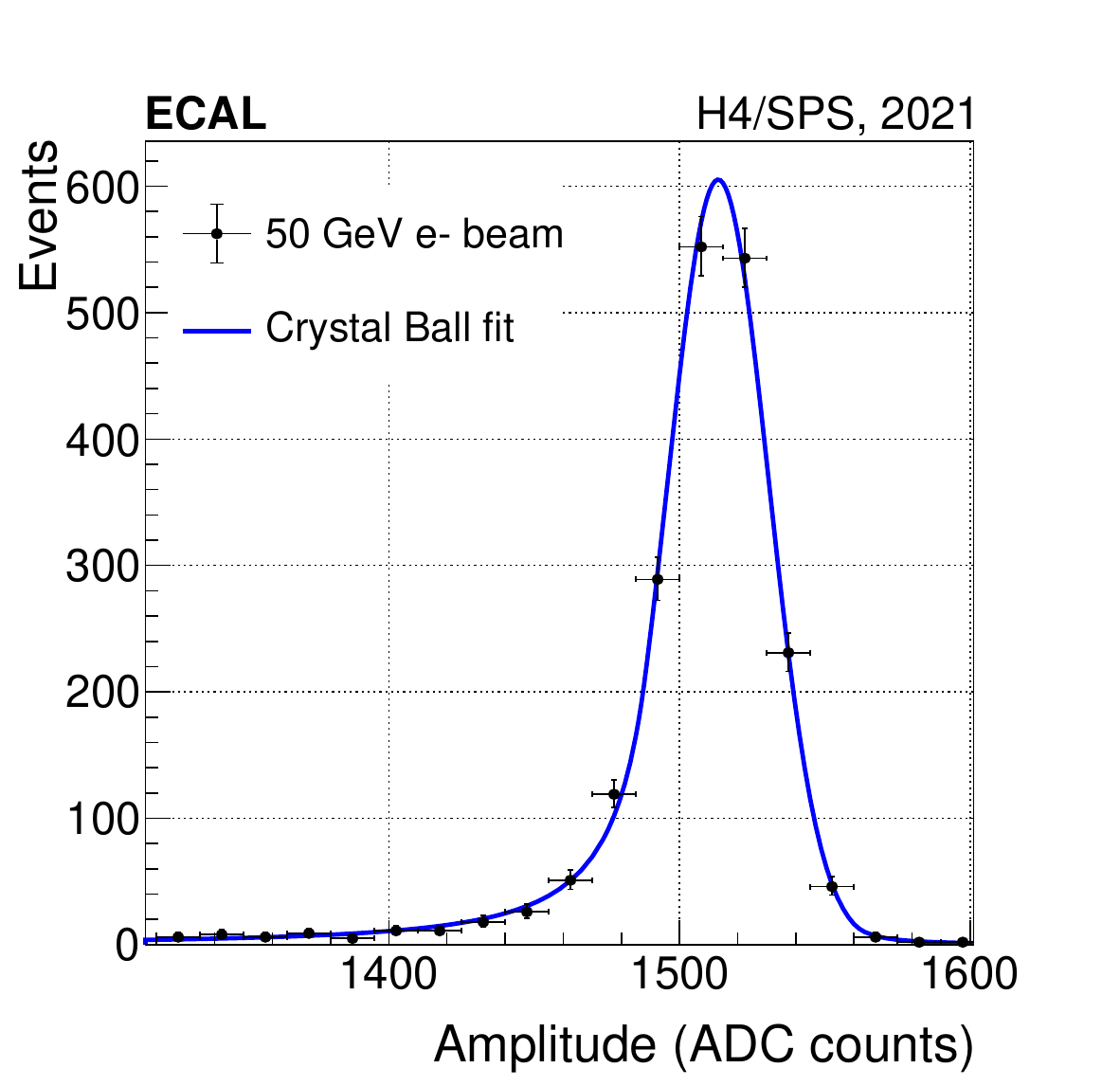}
        \caption{Example of the fit to the total amplitude distribution of the $3\times 3$ crystal matrix for a beam energy of 50\GeV. The blue line represents a Crystal Ball fit to the data, indicated by the black points.}
        \label{fig:tb:CB50GeV}
\end{figure}

The total amplitude in the $3\times 3$ matrix is defined as the sum of the signal amplitudes of the nine intercalibrated channels. For each beam energy, the total amplitude distribution is fitted with a double-sided Crystal Ball function~\cite{CrystalBall}, as reported in figure~\ref{fig:tb:CB50GeV}. Given the linear response of the ECAL to the energy of the impinging electromagnetic particles, the relative energy resolution is obtained from the ratio between the standard deviation and the mean of the Gaussian core of the Crystall Ball.

The function modelling the relative energy resolution $\sigma(E)/E$ versus the beam energy, $E$, is represented by the quadratic sum of three terms:
\begin{equation}
        \frac{\sigma(E)}{E} = \frac{N}{E} \oplus \frac{S}{\sqrt{E}} \oplus C,
        \label{eq:resole}
\end{equation}
where $N$, $S$ and $C$ are the noise, stochastic and constant terms respectively.
The stochastic term $S$ accounts for the statistical fluctuations of the electromagnetic shower in the crystal matrix and of the photons produced in each crystal. The constant term $C$ includes effects from the longitudinal leakage of the electromagnetic showers in the crystal matrix and from intercalibration uncertainties. In the fit of the relative energy resolution as a function of the beam energy, $S$ and $C$ are the free parameters. The term $N$ quantifies the average noise of the readout electronics of the $3\times 3$ matrix. As this term can be measured independently, it is fixed in the fit of the energy resolution to reduce parameter correlations.

For a single channel, the impact of the noise on the amplitude reconstruction has been estimated using a toy simulation. The noise, defined as the sample RMS, is measured on dedicated runs without beam. The full sample-to-sample covariance matrix is also measured.
The noise is then added to simulated waveforms of a fixed amplitude. The RMS of the reconstructed amplitude is found to be equal to that of the noise, increased by 2.5\%, independently of the amplitude.

\begin{figure}[t]
        \centering
        \includegraphics[width=.4\textwidth]{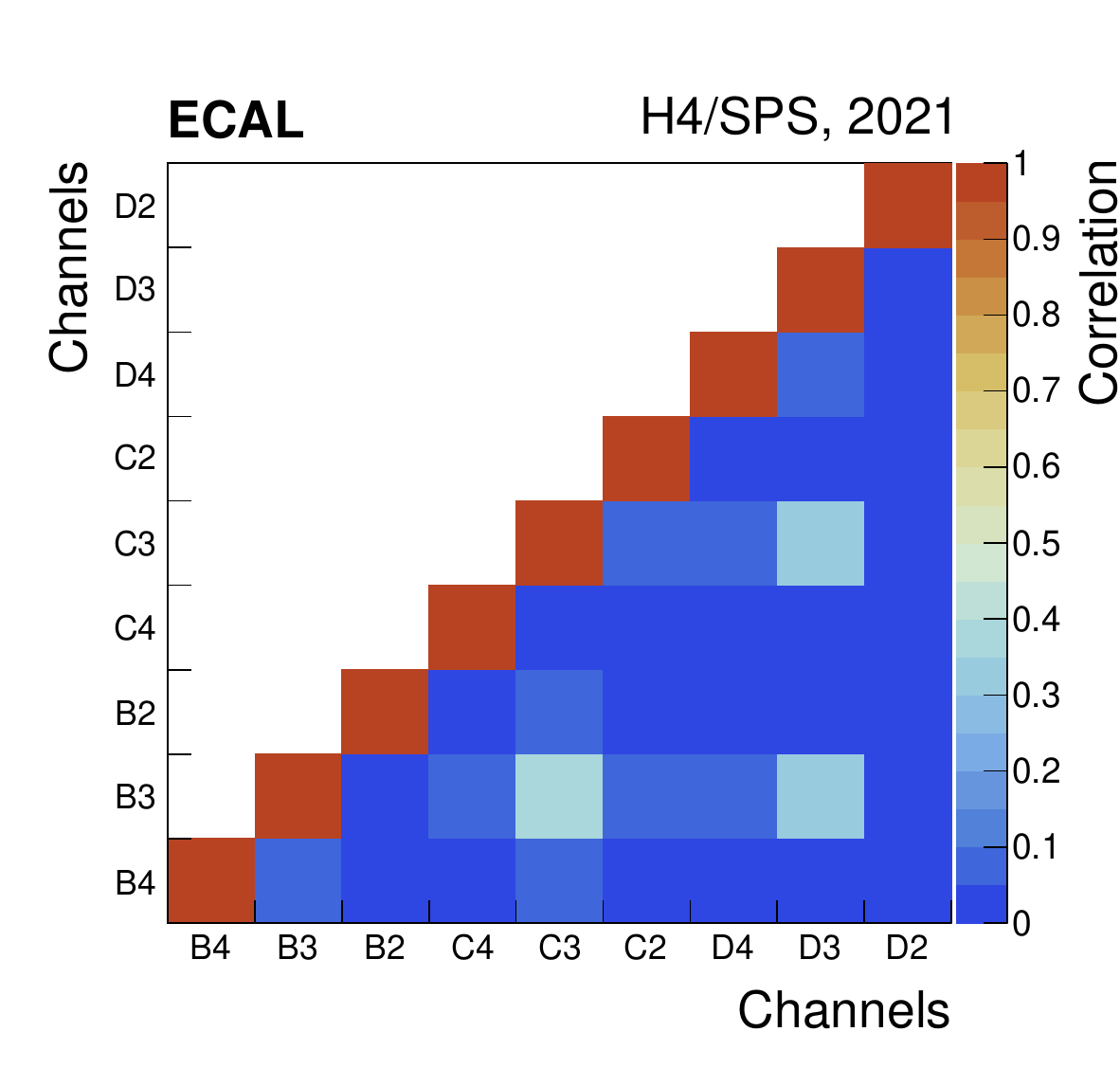}%
        \caption{Correlation matrix of the pedestal RMS for the 9 channels of the $3\times 3$ matrix $\mathbf{C}$.}
        \label{fig:tb:Cnoise}
\end{figure}

To estimate the term $N$ in Eq.~\ref{eq:resole} and to properly account for channel-to-channel correlations, the noise of the $3\times 3$ array is estimated from the noise of the individual channels as:
\begin{equation}
        N = 1.025 \cdot \sqrt{ \mathbf{B^T C B}},
\label{eq:noise}
\end{equation}
where $\mathbf{B}$ is the 9-dimension vector containing the average of the single channel noise, $\mathbf{C}$ is the $9\times 9$ correlation matrix of the noise between the different channels, and the factor 1.025 accounts for the 2.5\% noise increase described above. This correlation matrix is reported in figure~\ref{fig:tb:Cnoise}.

The values of the beam energy are corrected for estimated synchrotron radiation losses along the beamline, as provided in the beamline documentation and derived from a simulation of the beamline optics and components.

\begin{figure}[t]
        \centering
        \begin{minipage}{0.49\textwidth}
                \subfigure[]{\includegraphics[width=\textwidth]{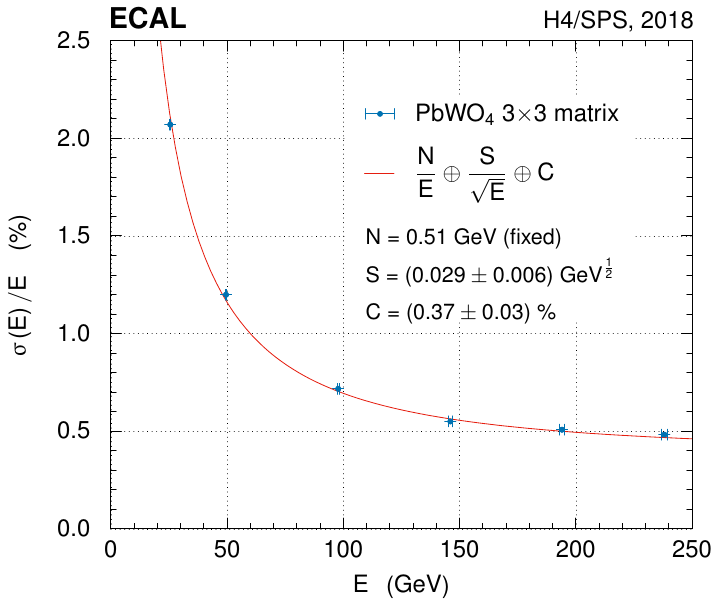}}
        \end{minipage}
        \begin{minipage}{0.49\textwidth}
                \subfigure[]{\includegraphics[width=\textwidth]{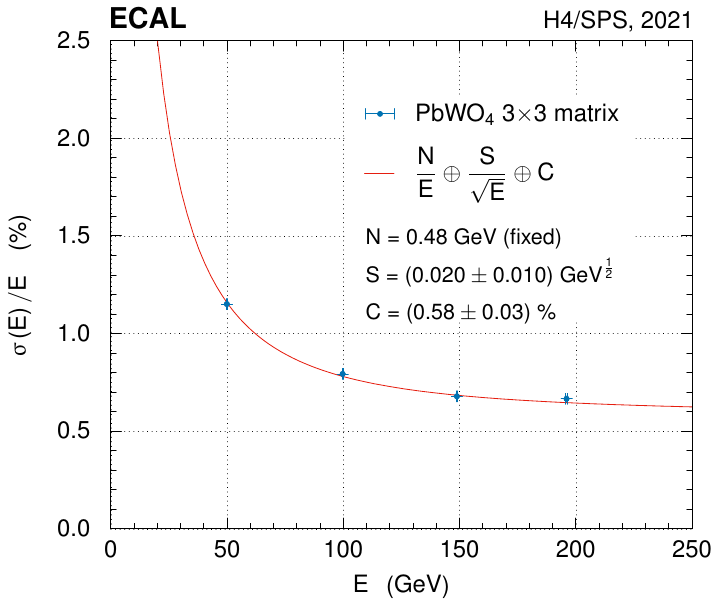}}
        \end{minipage}
        \caption{Relative energy resolution as a function of the beam energy obtained for the 2018 (a) and 2021 (b) campaigns. The energy deposits are measured in a $3\times 3$ matrix centred on the crystal impacted by the beam. Because of the significant correlation between the fit parameters, the noise term $N$ is fixed to the value measured in dedicated runs without beam. The error bars represent the statistical uncertainties.}
        \label{fig:tb:rese}
\end{figure}

The \PhaseTwo ECAL electronics targets a value for the constant term $C$ lower than 1\% at energies higher than 50\GeV. The energy resolution as function of the beam energy, measured in the two beam test campaigns, along with the fit of the S and C terms are  shown in figure~\ref{fig:tb:rese}.

The stochastic term is measured to be  $(0.029 \pm 0.006)\GeV^{1/2}$ and $(0.020 \pm 0.010)\GeV^{1/2}$ for the 2018 and 2021 campaigns, respectively, in agreement with beam test results of the legacy system~\cite{Adzic:2007mi}.
The constant term is measured to be $(0.37 \pm 0.03)\%$ and $(0.58 \pm 0.03)\%$ for the 2018 and 2021 campaigns, respectively. The difference between these two values is attributed to a misalignment of the beam with respect to the crystal matrix, which did not occur in the 2018 campaign. The constant term measured at 2018 beam test is slightly higher than the one measured in beam tests with the legacy system. This is assumed to be a consequence of the test matrix using crystals originating from the commissioning production batches, which have suboptimal characteristics compared to those used in ECAL. Additionally, the amount of material in front of the calorimeter was different in the two sets of measurements, and the 2018 and 2021 results have not been corrected for contributions from the energy spread of the beam. Overall, given these considerations, the energy resolution performance meets the requirements for the HL-LHC.

\begin{figure}[h!]
        \centering
        \begin{minipage}{0.51\textwidth}
                \subfigure[]{\includegraphics[width=\textwidth]{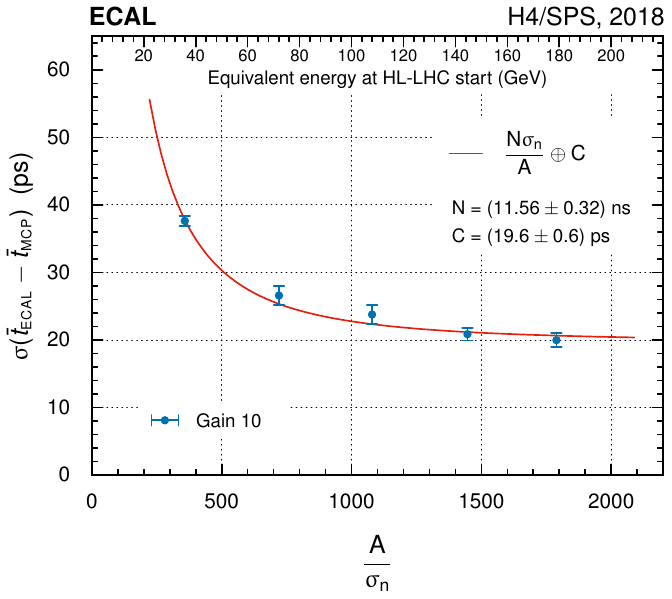}}
        \end{minipage}
        \begin{minipage}{0.49\textwidth}
                \subfigure[]{\includegraphics[width=\textwidth]{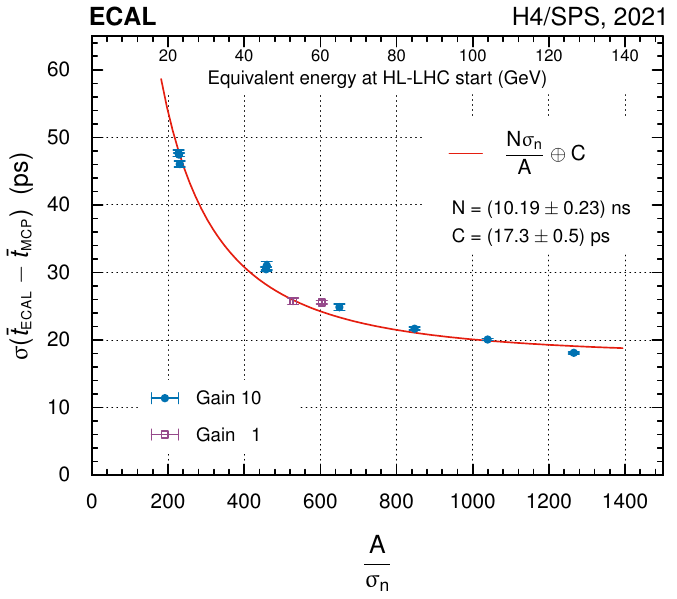}}
        \end{minipage}
        \begin{minipage}{0.49\textwidth}
                \subfigure[]{\includegraphics[width=\textwidth]{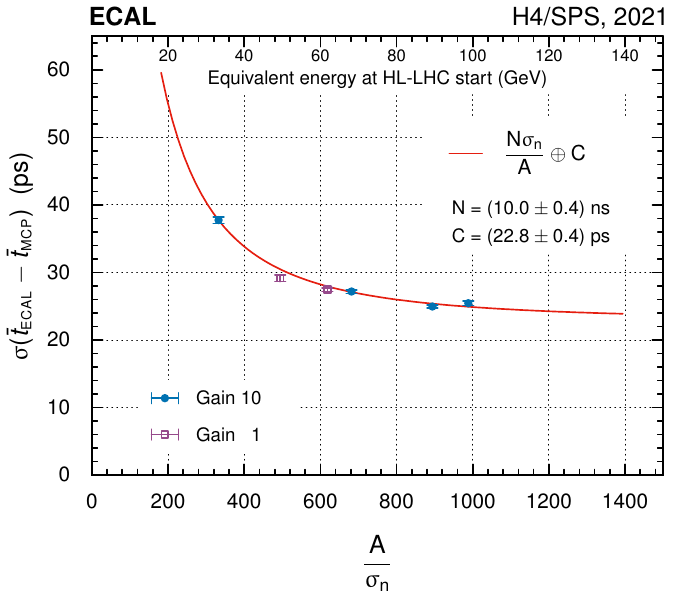}}
        \end{minipage}
        \caption{Time resolution as a function of the average reconstructed amplitude normalized by the noise RMS, $A/\sigma_{\text{n}}$, for a single crystal with the beam impinging within a $1\times 1\cm^2$ window, positioned to contain the maximum shower energy. Figures (a) and (b) show the resolution for the central channel of the ECAL matrix during the 2018 and 2021 beam tests, respectively, while (c) presents the resolution for a crystal adjacent to the central one. All plots present results after the subtraction of the MCP contribution to the resolution, as described in the text. The results for the 2021 beam test also include two points obtained at high energies, for which the electronics readout switches to gain 1. The error bars represent the statistical uncertainties. The red curve is a fit of Eq.~\ref{eq:rest} to the points in gain 10.}
        \label{fig:tb:rest}
\end{figure}

Using events where the beam hit the centre of a crystal, the single ECAL channel time resolution is extracted from the width of the $\bar t_{\textsc{ecal}} - \bar t_{\textsc{mcp}}$ distribution for electrons impinging within a window of $1\times 1\cm^2$ around the position of the maximum shower containment.
The time resolution for an individual channel is modelled as a function of the effective amplitude, i.e. the measured amplitude $A$ normalized to the channel average noise RMS $\sigma_n$. For any given channel, the quantities $A$ and $\sigma_n$ are derived independentely, with the amplitude $A$ obtained from the template fit to the digitized signal and $\sigma_n$ measured from the average of the pedestal RMS.

The analytic expression of the time resolution function, used to perform the fit to the data, is given by the quadratic sum of two terms:
\begin{equation}
        \sigma_t = \frac{N}{A/\sigma_n} \oplus C,
\label{eq:rest}
\end{equation}
where $N$ is a noise term and $C$ is the constant term. The contribution of the stochastic term is consistently found to be negligible in all beam test campaigns with legacy and \PhaseTwo electronics, and is therefore omitted from Eq.~(\ref{eq:rest}). The results are shown in figure~\ref{fig:tb:rest}, where each point on the $x$-axis is obtained from the peak of a Crystal Ball function fitted to the distribution of $A/\sigma_n$.

The MCP time resolution, $\sigma_t^{\textsc{mcp}}$, is subtracted in quadrature from each of the measured $\sigma(\bar t_{\textsc{ecal}}-\bar t_{\textsc{mcp}})$ values, before fitting the data. Assuming the MCPs are identical, $\sigma_t^{\textsc{mcp}}$ is measured as the spread of the time difference between the two MCPs, divided by $\sqrt{2}$, considering all the events contributing to each $\sigma(\bar t_{\textsc{ecal}}-\bar t_{\textsc{mcp}})$ measurement, separately. The value of $\sigma_t^{\textsc{mcp}}$ unfolded from each point is basically independent of the beam energy and amounts to approximately 13\ps.

During the 2021 beam test, the time resolution was measured both for low and high electronics gain. Only runs with beam energies exceeding 150\GeV have a sufficient amount of data in the low gain mode, with measurements at 200 and 250\GeV.

The results fulfil the design goals of having a resolution of 30\ps or better at the beginning of HL-LHC for energies greater than 50\GeV.

\section{Summary}

Results have been presented from studies of the performance of first prototypes of the CMS ECAL barrel readout electronics that will replace the existing system during the future high luminosity operation of the LHC. The measurements were made with high energy electrons and pions in the H4 and H2 beam lines of the CERN SPS, in separate campaigns in 2018 and 2021.

The new amplifier bandwidth coupled with the high sampling frequency provides improved discrimination between signals from electromagnetic showers and from direct ionization of the avalanche photo-diodes. The performance of the \PhaseTwo readout electronics has been characterized in terms of time and energy resolutions, measured on a $5\times 5$ matrix of lead tungstate crystals. Two multi channel plate detectors have provided references for time measurements.

The results for the energy resolution, primarily driven by the design of the detector, in absence of photodetector ageing and crystal irradiation, are compatible with previous publications with the legacy electronics. In both campaigns the constant term of the energy resolution is measured to be better than 0.6\% and the time resolutions for electrons with energies above 50\GeV is measured to be better than 30\ps, fulfilling the design requrements.

\section{Acknowledgments}

The authors would like to thank the support of CERN for the excellent beam test facilities and operations.

We warmly thank M. and A. Barnyakov for providing us with and for valuable information on the Ekran FEP photomultipliers used in this study.

We would also like to thank the technicians and engineers, from CERN and collaborating Institutes, who worked on the preparation of the crystals and its setting up and operation in the beam test, as well as those setting up the equipment in the beam area.

\bibliography{auto_generated}

\cleardoublepage \appendix\section{The CMS electromagnetic calorimeter group \label{app:collab}}\begin{sloppypar}\hyphenpenalty=5000\widowpenalty=500\clubpenalty=5000\cmsinstitute{Centro Brasileiro de Pesquisas Fisicas, Rio de Janeiro, Brazil}
{\tolerance=6000
G.A.~Alves\cmsorcid{0000-0002-8369-1446},
P.~Rebello~Teles\cmsorcid{0000-0001-9029-8506},
M.G. Ferreira Siqueira Amaral Gomes\cmsorcid{0000-0003-0483-0215}
\par}
\cmsinstitute{Institute of High Energy Physics, Beijing, China}
{\tolerance=6000
T.~Cao\cmsorcid{0009-0001-0026-6154},
M.~Chen\cmsAuthorMark{1}\cmsorcid{0000-0003-0489-9669},
S.~Song\cmsAuthorMark{1}\cmsorcid{0009-0005-5140-2071},
J.~Tao\cmsorcid{0000-0003-2006-3490},
C.~Wang\cmsAuthorMark{1}\cmsorcid{0009-0009-7236-5563},
J.~Wang\cmsorcid{0000-0002-3103-1083},
J.~Zhang\cmsorcid{0009-0007-2515-1808}
% ...
\par}
\cmsinstitute{Authors affiliated with an international laboratory covered by a cooperation agreement with CERN}
{\tolerance=6000
V.~Matveev\cmsAuthorMark{4}\cmsorcid{0000-0002-2745-5908}
\par}
\vskip\cmsinstskip
$^{1}$Also at University of Chinese Academy of Sciences, Beijing, China\\
$^{2}$Now at CERN, European Organization for Nuclear Research, Geneva, Switzerland\\
$^{3}$Now at IHEP, Beijing, China\\
$^{4}$Also at another institute formerly covered by a cooperation agreement with CERN\\
\end{sloppypar}
\end{document}